# Synergistic Effects of Natural Biosurfactant and Metal Oxides Modification on PVDF Nanofiber Filters for Efficient Microplastic and Oil Removal

Aleksander de Rosset[1], Rafael Torres-Mendieta[2], Grzegorz Pasternak[1], Fatma Yalcinkaya[*,2]

[1]Laboratory of Microbial Electrochemical Systems, Department of Process Engineering and Technology of Polymer and Carbon Materials, Faculty of Chemistry, Wroclaw University of Science and Technology, Wroclaw 50-370, Poland.

[2]Institute for Nanomaterials, Advanced Technologies and Innovation, Technical University of Liberec, Studentská 1402/2, 461 17 Liberec, Czech Republic.

[*]fatma.yalcinkaya@tul.cz

**Abstract:**

The removal of microplastics and oil from oil-water emulsions presents significant challenges in membrane technology due to issues with low permeability, rejection rates, and membrane fouling. This study focuses on enhancing nanofibrous composite membranes to effectively separate microplastic contaminants (0.5µm) and oil-water emulsions in wastewater. Polyvinylidene fluoride (PVDF) polymeric nanofibers were produced using a needle-free electrospinning technique and attached to a nonwoven surface through lamination. The membranes were modified with alkaline treatment, biosurfactant (BS), $TiO_2$, and CuO particles to prevent fouling and improve separation efficiency. The modified membranes demonstrated exceptional water permeability, with BS-modified membranes reaching above 9000 $Lm^{-2}h^{-1}bar^{-1}$ for microplastic separation. However, BS modifications led to reduced water permeability during oil-water emulsion treatment. $TiO_2$ and CuO further enhanced permeability and reduced fouling. The $TiO_2$-modified membranes exhibited superior performance in oil-water emulsion separation, maintaining high oil rejection rates (~95%) and antifouling properties. The maximum microplastic and oil rejection rates were of 99.99% and 95.30%, respectively. This study illustrates the successful modification of membrane surfaces to improve the separation of microplastics and oil-water emulsions, offering significant advancements in wastewater treatment technology.

## 1. Introduction:

The escalation of water resource contamination has become a pressing concern due to the rapid population growth along with our society's intense industrialization, particularly within the food, oil/gas, petrochemical, automotive, machinery, and pharmaceutical sectors[1]. This surge in industrial activities has led to a significant increase in the production of various types of water pollution, with oil and plastic contamination, posing the most alarming threats. The reintroduction of these pollutants into our societal life cycle has substantial consequences, including biodiversity decline due to exposure to toxic hydrocarbons from oil[2], health risks associated with hazardous microplastic exposure resulting from plastic's structural degradation rendered by biological, physical, and chemical processes[3]. Every year, large quantities of pollutants mainly including heavy metals, oily wastewater and microplastics (MPs) waste are released, resulting in pollution of the water resources. This results in serious threads the ecosystems, aquatic environment, and human health. In particular, MPs and emulsified oil in water have the greatest impact on the environment. This scenario translates into a massive economic impact, affecting industries reliant on water resources, leading to financial losses and heightened water treatment costs[4].

Several new technologies for removing MPs from aquatic environments have been developed in recent years[5]. Recently, a durable and compressive sponge prepared from chitin and graphene oxide was proposed for effective adsorption of various types of MPs from water [6]. As a result, the adsorption capacity of polystyrene MPs was almost 90%. Subsequently, MPs could undergone a biodegradation. Such an approach would thus allow simultaneously capture and degrade MPs. Although most of MPs are generally considered as non-biodegradable, recent studies showed, that microorganisms are capable of colonizing MPs surfaces and degrading them [7]. Similarly, effective technologies are needed for the removal of emulsified oil from water. Physical methods mainly used as primary treatment of oily wastewater include gravity separation [8] , hydrocyclones separation [9] and flotation [10], while chemical methods involve the use of various types of deemulsifiers and biodeemulsifiers [11,12]. However, the use of physical or chemical methods for separating emulsified oil not always allow comply with respective pollution standards. In such case, it is necessary to subsequently treat oily wastewater using advanced oxidation [13], filtration [14,15], or biodegradation processes in bioreactors [16].

Among various water treatment strategies, polymeric membrane filtration has emerged as the golden standard due to its effectiveness, selectivity, efficiency, consistency, environmental friendliness, scalability, and regulatory compliance [17,18]. However, despite these benefits, the technology faces significant challenges, particularly in fouling, selectivity, specificity in targeting contaminants, ensuring chemical compatibility and long-term durability, environmental friendliness, and adaptability to emerging contaminants [19–21]. Traditionally, addressing these challenges involves enhancing membrane properties through doping, typically with metal oxides like $TiO_2$ or CuO. These metal oxides contribute to fouling resistance, superficial chemistry modification for improved selectivity and specificity, and compatibility with a broader range of water chemistries, making the membranes more sturdy in various environmental conditions [22,23].

While doping polymeric membranes with metal oxides shows promise, it is essential to note that doping strategies are often not universal, especially when forming membranes with chemically robust polymers like polyvinylidene fluoride (PVDF), which requires the use of functional molecules to bind metal oxides only on the membrane surface, maximizing the properties of these dopants and avoiding atom wastefulness through doping within the polymer[24,25]. From all possible functional molecules allowing an appropriate metal oxide functionalization on the PVDF surface, in the current manuscript, we are exploring for the first time the use of rhamnolipids, glycolipid biosurfactants produced by many bacterial species, predominantly *Pseudomonas aeruginosa*[26]. These biosurfactants have a great potential for removing oil and microplastics from wastewater due to their non-toxicity, high biodegradability, and high capabilities in reducing surface tension. In our recent study, this modification approach has become successful in preventing the undesirable biofouling of PVDF membranes which were used for proton transport across the electrodes in microbial fuel cell system [27]. When used to dope PVDF nanofibrous membrane with metal oxides, they can reduce liquid surface tension, enhance microplastic adsorption, and improve filtration efficiency, anti-fouling, and oil-water separation facilitated by the metal oxides. Although these amphiphilic biosurfactants have been employed to bind metal oxides to polymeric membranes for various applications [28–30], to the best of our knowledge, no attempt has been made to use them in the metal oxide doping of PVDF nanofibrous membrane, which chemical resistance, thermal stability, and mechanical strength making them one of the top candidates towards technology transfer [31,32]. Thus, the pioneering use of rhamnolipids in the metal oxide doping of these tough nanofibers holds great promise in contributing to the ongoing efforts to mitigate the escalating depletion of clean water.

## 2. Materials and Methods

### 2.1. Preparation of PVDF Nanofibrous Membrane

The PVDF (Kynar, FR) nanofibers were produced using a needle-free electrospinning system (Nanospider NS 8S1600U, Elmarco, CZ), employing a solution containing 13 wt. % PVDF in N, N-dimethylformamide (DMF, Penta s.r.o., CZ). The resulting nanofibers were gathered on silicon paper and subjected to lamination onto a spunbond polyethylene terephthalate (PET) nonwoven fabric (100 g/m$^2$, Mogul Nonwovens, TUR) using a co-polyamide adhesive layer (Protechnic, FR). A heat press (HLV 150, Pracovní stroje Teplice s.r.o., CZ) facilitated the lamination process (130ºC, 50 kN) for 3 minutes. The resulting composite material demonstrates resilient tensile strength, capable of withstanding external forces during membrane separation processes [18,19,33].

### 2.2. Surface Modification of the Membrane

The membrane surface underwent modification at various stages. Initially, the composite PVDF nanofibrous membrane was treated with a 1.8 M potassium hydroxide (KOH, Fluka, CZ) solution in isopropanol under ambient conditions for 15 minutes. As discussed in previous studies[23,34–36], this chemical treatment facilitates the dehydrofluorination of the PVDF membrane, introducing OH groups that allow for the attachment of diverse functionalities, including hydroxyl functional groups, esters, ethers, amines, aldehydes, nanoparticles, and metal oxides (the resulting membrane was labeled as PVDF-OH). In the second stage, the PVDF-OH membrane underwent modification using a biosurfactant (specifically, Rhamnolipids 3-O-(α-L-rhamnopyranosyl-(1-2)-α-L-rhamnopyranosyl)-3-hydroxydecanoic acid) dispersed in distilled water at two concentrations: 0.5 g/L and 2 g/L. The procedure involved immersing the membranes in the solution for 1 day, enabling the formation of the self-assembled monolayer, which resulted in the membranes labeled as PVDF-OH/BS-0.5 and PVDF-OH/BS-2 for the lower and higher concentrations, respectively.

The employed rhamnolipid featured a glycosyl head group, specifically a rhamnose moiety, along with a 3-(hydroxyalkanoyloxy) alkanoic acid (HAA) fatty acid tail, represented by 3-hydroxydecanoic acid. These structural characteristics found in the employed rhamnolipid facilitate the attachment of the biosurfactant to the PVDF-OH membrane due to their interaction with the membrane hydroxyl groups[37]. The interaction between BS and PVDF-OH has been explained previously [27]. The interaction between the hydroxylated PVDF and biosurfactant is enhanced by the presence of mutually supporting binding sites, which facilitate electron transfer between the two entities.

In the third phase, the PVDF-OH/BS membrane underwent further modification by introducing titanium dioxide (TiO$_2$, Sigma Aldrich, 20nm) nanoparticles and copper oxide (CuO, 79.55 g/mol, with a size distribution of ~1μm, Penta s.r.o.) microparticles. The modification process involved immersing the membranes in a 2 wt. % aqueous metal oxide suspension for three days. These nano- and microparticles could interact with the hydroxyl groups on the polymer surface through mechanisms like physical adsorption and other non-covalent interactions. Subsequently, the adapted membranes underwent a 10-minute ultrasonic bath treatment to remove any excess and loosely attached nano and microparticles from the membrane surface. The labels for these modified membranes are listed in **Table 1**.

**Table 1.** List of samples' labels

| Abbreviation | Chemical Treatment | BS amount | TiO$_2$ | CuO |
|---|---|---|---|---|
| PVDF | - | - | - | - |
| PVDF-OH | In KOH | - | - | - |
| PVDF-OH/BS-0.5 | In KOH | 0.5g/L | - | - |
| PVDF-OH/BS-2 | In KOH | 2.0g/L | - | - |
| PVDF-OH/BS-0.5/TiO$_2$ | In KOH | 0.5g/L | 2 wt. % | - |
| PVDF-OH/BS-2/TiO$_2$ | In KOH | 2.0g/L | 2 wt. % | - |
| PVDF-OH/BS-0.5/CuO | In KOH | 0.5g/L | - | 2 wt. % |
| PVDF-OH/BS-2/CuO | In KOH | 2.0g/L | - | 2 wt. % |

## 2.3. Filtration Test

Two separate filtration experiments were carried out. The first test targeted the elimination of microplastics, whereas the subsequent one aimed to segregate oil-water emulsions. The experiments utilized a laboratory-scale dead-end filtration setup, functioning under a pressure of 0.02 bar. For the initial filtration trial, distilled water was used with all the membranes, and the membrane flux and permeability were calculated using equations (1) and (2).

$$j = \frac{L}{At} \quad (1)$$

$$k = \frac{j}{P} \quad (2)$$

where, $j$ is membrane flux (L/(m²h), $L$ is the volume of permeate in a liter (L), $A$ is the active membrane area (m²), $t$ is the filtration time (h), $k$ is the membrane permeability (L/(m²hbar)), and $P$ is the transmembrane pressure (bar).

The calculation of rejection ratio ($R$ (%)) was done through the equation (3);

$$R = \left(1 - \frac{C_f}{C_i}\right) x\ 100\ (\%) \quad (3)$$

where $C_i$ represents the initial concentration of the contaminant in the feed, and $C_f$ denotes the concentration of the contaminant in the permeate liquid once the separation cycles have been completed.

Membrane anti-fouling has been measured by using flux recovery rate (FRR) and total decline in flux rate (Rt) were calculated from equations (4) and (5) [38,39].

$$FRR = \frac{Jw_2}{Jw_1} x\ 100\ (\%) \quad (4)$$

$$Rt = \frac{J_i - J_f}{J_i} x\ 100\ (\%) \quad (5)$$

where, Jw$_1$ is pure water flux, Jw$_2$ is pure water flux after oil-water cycle, J$_i$ and J$_f$ are the initial and final flux before and after fouling with oil-water emulsion, respectively.

*Microplastic Removal Test:* The efficacy of the modified PVDF membranes in removing microplastics was assessed compared to the original, unaltered membrane. To achieve this, 0.5 μm polydispersed polystyrene (PS) microparticles were employed. Specifically, 5 μL of PS particles were added to 1 L of distilled water and agitated for 10 minutes. The filtration test extended over a period during which the membranes' microplastic removal efficiency was evaluated by analyzing the turbidity test carried out using a turbidity meter, measured in NTU units, to further confirm the effectiveness of particle removal.

*Oil Removal Test:* The oil separation test involved combining 5 g of sunflower oil, 100 ml of distilled water, and 0.1 ml of Triton X-100 (laboratory-grade, Sigma-Aldrich). The mixture underwent overnight stirring at 600 rpm until a consistent emulsion was attained. Using an optical microscope (Axio Imager M2, Carl Zeiss), emulsion uniformity and drop sizes were measured, revealing an average oil droplet diameter of 6 ± 3 μm for kitchen oil. Subsequently, the emulsion was filtrated with the produced membranes 10 times (10 cycles) without changing them to evaluate the sample's resistance to fouling. The amount of organically bound carbon in the oil/water emulsion was quantified by determining the chemical oxygen demand (COD) using the potassium dichromate oxidation method, according to the manufacturer's instructions (Hach, USA). The rejection of pollutants was determined by employing Equation (3).

## 2.4. Membrane Characterization

The surface morphology of the prepared samples was investigated using a Zeiss Ultra Plus scanning electron microscope (SEM) fitted with an Oxford X-Max 20 energy-dispersive (EDS) detector, and the Image-J software was employed to determine fiber diameter. A Nicolet iZ10 Fourier Transform Infrared Spectroscope (FTIR; Thermo Scientific, CZ) was utilized to assess alterations in the chemical structure of the membrane surface due to surface modification, and adding to the SEM-EDS examination, a thorough analysis of the membranes' surface characteristics was conducted by measuring its roughness using an Atomic Force Microscope (AFM; JPK Nanowizard III, Bruker Corporation) equipped with a cantilever (PPP-CONTSCR, NANOSENSORSTM).

The membrane pore size was determined through the bubble point method using a pore size analyzer (Porometer 3G, Anton Paar GmbH). A minimum of three measurements were taken, and the average pore size was calculated, adhering to the standards outlined in ASTM F316-03 (Standard Test Methods for Pore Size Characteristics of Membrane Filters by Bubble Point and Mean Flow Pore Size). Water contact angle measurements of the membranes were performed using a Drop Shape Analyser DS4 (Krüss GmbH, Hamburg, Germany) with distilled water (surface tension: 72.0 mN/m). The contact angle measurements followed ASTM D 5725-99 (Standard Test Methods for Surface Wettability and Absorbency of Sheeted Materials Using an Automated Contact Angle Tester). Each measurement was taken from both the right and left sides of the droplet, and averages were calculated. A minimum of five measurements were obtained from various locations on the samples, and an overall average was derived.

Furthermore, the NPs leaching into the permeate post-separation was determined through single-particle inductively coupled plasma–mass spectrometry (spICP–MS) using a spectrometer (NexION 3000D, Perkin Elmer) with a detection limit of 0.5 ng/L.

## 3. Results and Discussion

### 3.1. Structure and Surface Morphology of the Membrane

The filtration efficiency of the nanofibrous membrane is significantly influenced by its structural characteristics. **In Fig. 1**, the membrane modification process is illustrated. The initial stage involves an elimination reaction of PVDF in a KOH rich environment, resulting in the removal of hydrogen (H) and fluorine (F) elements from the polymer chain, forming a C=C double bond[40]. Subsequently, when the modified polymer chain comes into contact with a solution rich in hydroxide ions, it undergoes a series of reactions that lead to the incorporation of hydroxyl groups as outlined below [38]:

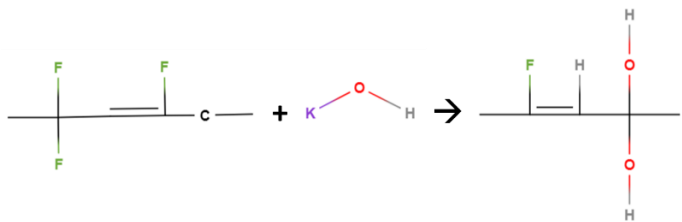

**Fig. 1.** Modification of PVDF membranes with KOH (dehydrofluorination).

The second stage includes the surface modification of the dehydrofluorinated PVDF with biosurfactant Rhamnolipids, as previously described [27].

The presence of a defect-free nanofibrous layer is of utmost importance as it significantly influences the membrane rejection efficiency. During the lamination process, significant attention was given to controlling temperature and applied force in order to preserve the integrity of the composite membrane structure and prevent any surface damage. As depicted in **Fig. 2**, the surface of the nanofibrous membranes remained undamaged following the lamination and modification procedures.

**Fig. 2** depicts the surface morphology and chemical composition of the samples, as determined from the EDS spectra. Membrane without NPs decoration shows smooth fiber surface with similar fiber diameter (Fig.2 (a-d)) The EDS analysis of $TiO_2$ nanoparticles incorporated PVDF membranes reveals the presence of only Ti, O, C, and F elements (**Fig. 2 e and f**). The percentage of Ti and O elements increased as the quantity of biosurfactant used for modification increased, providing evidence that the $TiO_2$ nanoparticles effectively bind to the biosurfactant. In the case of $TiO_2$, a trace amount of Al, less than 0.1% by mass, is observable (**Fig. 2e**), likely originating from the aluminum support. Likewise, the analysis of CuO microparticles revealed the presence of Cu, C, and F elements (**Fig. 2 g and h**). The percentage of Cu on the membrane was relatively minimal, to the extent that O elements could not be detected using SEM-EDS due to their low concentration. This observed scarcity of CuO may be attributed to the limited affinity of the biosurfactant for the larger size of the CuO particles.

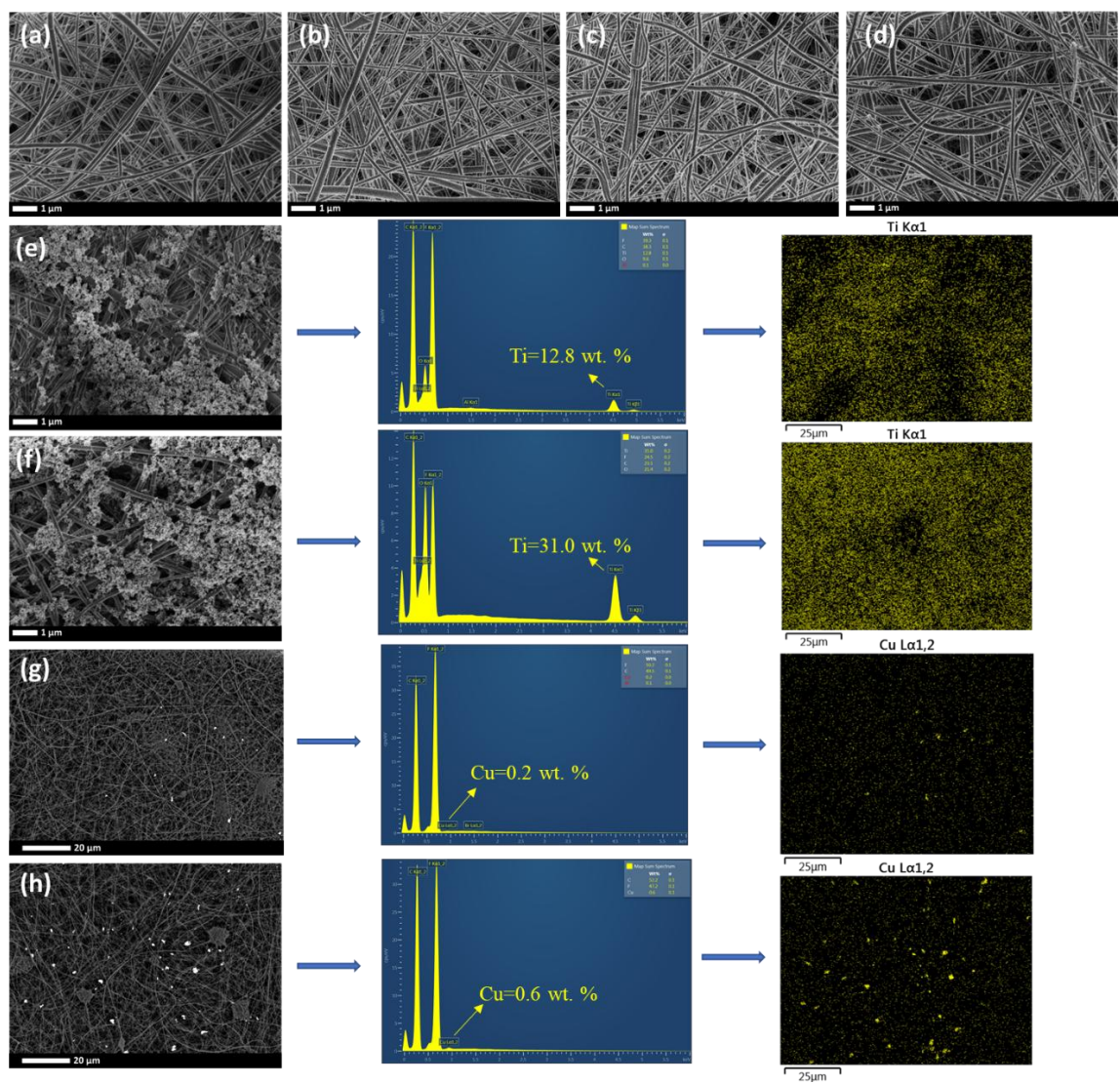

**Fig. 2.** SEM-EDS images of (a)pristine PVDF, (b) PVDF-OH, (c) PVDF-OH/BS-0.5, (d) PVDF-OH/BS-2, (e) PVDF-OH/BS-0.5/TiO2, (f) PVDF-OH/BS-2/TiO2, (g) PVDF-OH/BS-0.5/CuO, (h) PVDF-OH/BS-2/CuO.

The membrane's maximum, minimum, and average pore sizes were assessed both before and after undergoing a modification process with a biosurfactant, as depicted in **Fig. 3**. The findings revealed that the mean pore size experienced a minor increase following alkaline modification, though the change was not particularly substantial. The data regarding pore size are essential for understanding the membrane's ability to reject microplastics and emulsified oil.

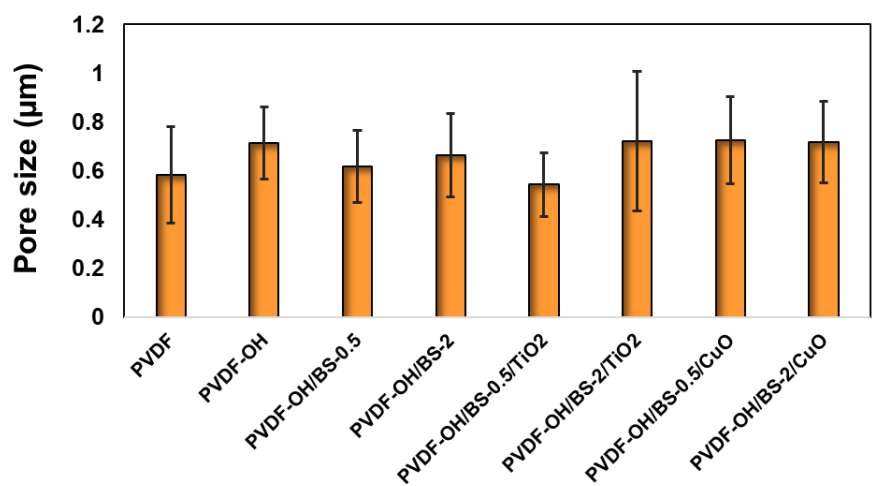

**Fig. 3.** Pore size of the modified and pristine membranes

FTIR results show the chemical composition of the unmodified and modified membranes (**Fig. 4**). The prominence of the -CH$_2$ rocking vibration peak of PVDF between 730-740 cm$^{-1}$ increases in the modified samples, whereas the opposite trend is observed for the bending vibration of -CF$_2$, which is noted at 612 cm$^{-1}$. The modification might affect the vibrational behavior of the CF$_2$ groups, potentially increasing the intensity of the rocking vibration peak. Each material exhibited characteristic C-H stretching vibrations around ~2859 and 2989 cm$^{-1}$. The peak at 1718 cm$^{-1}$ for the BS-modified PVDF polymer indicates the presence of carbonyl (C=O) stretching vibrations, which can often be associated with the presence of functional groups such as esters.

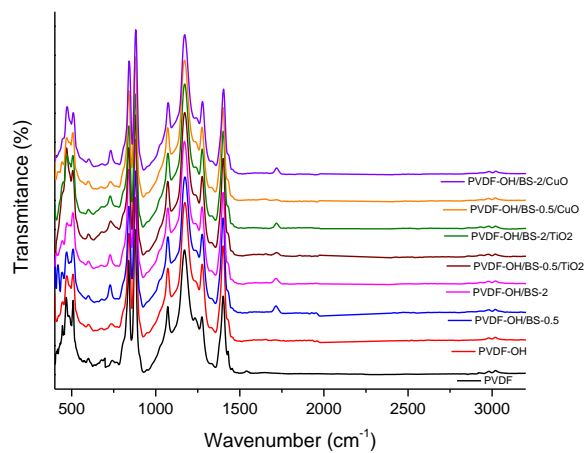

**Fig.4.** FTIR results of pristine and modified PVDF nanofibrous membranes.

The water contact angle (WCA) measurements (**Fig. 5**) conducted on both the pristine and modified membranes indicate that the alkaline treatment alone did not sufficiently enhance the hydrophilicity of the membrane. In contrast, the dried PVDF-OH membrane exhibited a hydrophobic nature. While the PVDF-OH membrane displayed hydrophilic characteristics in water, the hydrophilicity vanished after the membrane was dried. This change can be attributed to the flip-flop mechanism, a common occurrence in surface functionalities that are attached to polymer membranes [41]. When the external environment is hydrophobic, the hydrophilic -OH groups on the membrane's surface may reorient from the exterior to the interior of the polymer surface, resulting in the loss of hydrophilic hydroxyl groups on the surface. Consequently, the membrane's surface becomes hydrophobic. This underscores the importance of maintaining membranes in an aqueous solution promptly after hydrophilization to preserve their desired properties. Similar results has been reported previously [38].

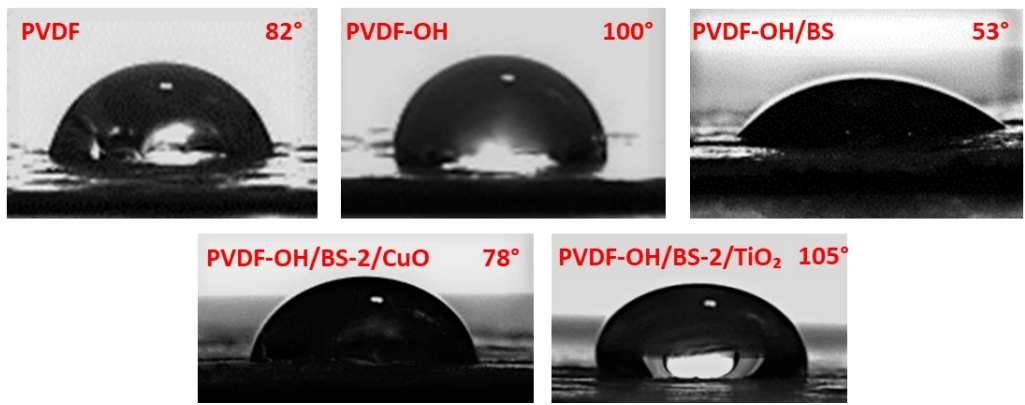

**Fig. 5.** Water contact angle of pristine and selected modified PVDF membranes.

The results from the WCA analysis demonstrated that the addition of BS can improve the hydrophilicity of the membrane, even when it is in a dry state. The membrane can be stored in air without experiencing changes in its wettability qualities. It is well-established that CuO increases the hydrophilicity of nanofibers [42,43]. Since the amount of attached CuO is not much on the membrane (Fig. 2 (g) and (h)), the membrane hydrophilicity decreased slightly compared to pristine PVDF membrane. On the other hand, $TiO_2$ decorated membranes exhibited hydrophobic properties. The affinity of $TiO_2$ for water depends on its crystal phase. Rutile $TiO_2$ exhibits lower affinity for water, whereas anatase $TiO_2$ demonstrates a higher affinity for water [44,45]. Surface roughness also influences the WCA of the membrane. The nanoparticles can create a rough surface that traps air, reducing the contact between the water and the solid surface. The AFM result (**Fig. 6**) confirmed that $TiO_2$-modified membranes have a highly rough surface.

Although the roughness values measured across the entire membranes do not seem to follow any trend regarding the performed modification (**Fig. 6**), examining the roughness on individual fibers' surfaces reveals a strong correlation with contact angle measurements and FTIR data. Nanofibers in pristine PVDF and PVDF-OH exhibit low roughness, but their lack of hydrophilic functional groups results in poor hydrophilicity, which can hinder efficient water permeability. Upon adding BS (samples PVDF-OH/BS-0.5 and PVDF-OH/BS-2), the fibers' roughness does not significantly change, but the incorporation of hydrophilic functional groups, including C=O [46], increases the membrane's hydrophilicity, which can potentially enhance permeability. In contrast, the addition of $TiO_2$ significantly increases the nanofibers' roughness, which reduces the contact area between the membrane surface and water droplets. This increases the effective surface tension, leading to hydrophobicity [47], as shown in **Fig. 5**. While this is not desirable for water permeation, it can improve rejection rates when recovering water from oily emulsions where oil contacts the membranes as droplets [23] or microparticle-polluted water [48], also displaying a phase separation. CuO addition, on the other hand, does not significantly alter the nanofibers' roughness, preserving the membrane's hydrophilicity to some extent. However, as observed in **Fig. 5**, the interaction between C=O and CuO seems to affect the overall membrane hydrophilicity. This might be due to CuO hindering the proper exposure of C=O groups [49].

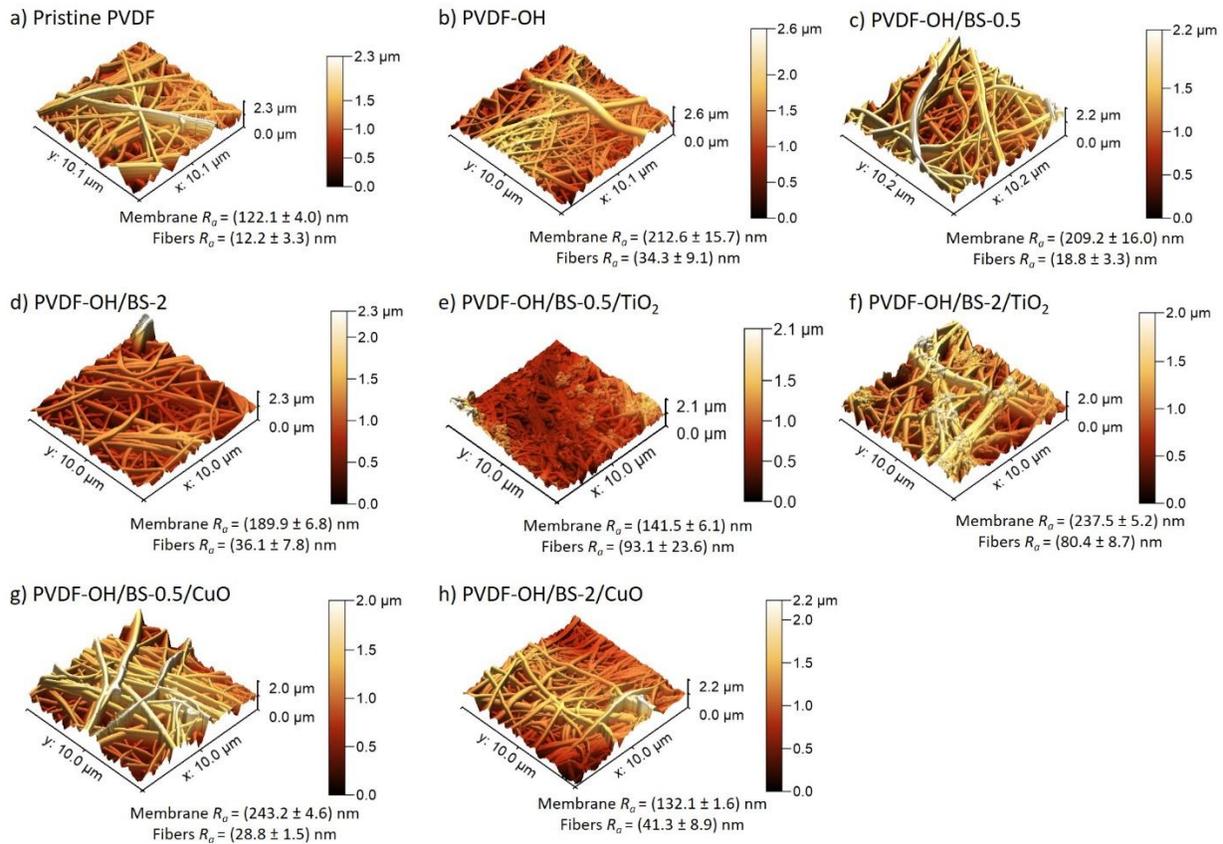

**Fig. 6.** AFM images of all samples including roughness measurements across the entire membranes and of individual fibers within the membranes.

### 3.2. Filtration of microplastics and emulsified-oil by PVDF nanofibrous membranes

The pristine and modified PVDF membranes were evaluated for their effectiveness in removing microplastics from water and oil from oil-water emulsions. The filtration process was conducted under a low pressure of (0.02 bars).

### 3.2.1. Microparticle Separation

The microplastic removal efficiency and membrane permeability has been given in **Fig.7**. Particle rejection was measured using a turbidity tester.

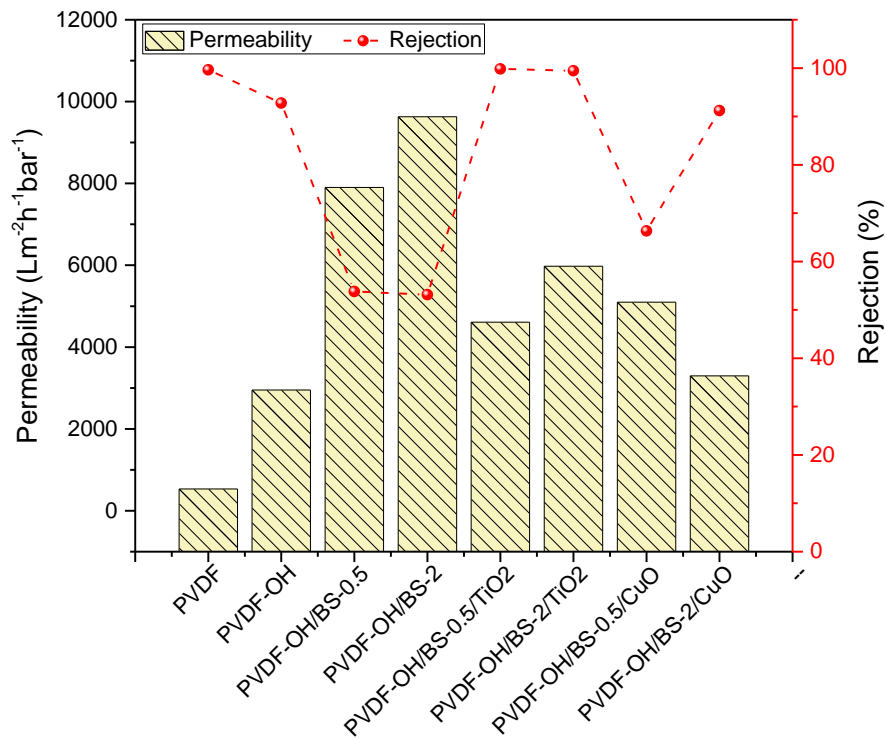

**Fig. 7.** Microparticle separation performance of membranes

The particle separation test demonstrated a significant enhancement in water permeability when the membranes were exposed to a higher concentration of biosurfactants. On the other hand, the modification with biosurfactant resulted in a significant decrease in particle rejection. The particle rejection test was done by using turbidity test. The reason could be due to release of BS to water which might influence the turbidity level of the water. Kalcikova et. al [50] studied the absorption kinetics of $nTiO_2$ on the polyethylene MPs. The experimental results indicated that $nTiO_2$ was rapidly adsorbed onto the surface of MPs, confirming the strong binding of $nTiO_2$ to the MPs' surface. Similarly, in our study, the membranes with $TiO_2$ demonstrated extremely high rejection rates for the MPs. The reason could be the attraction between the oppositely charged $TiO_2$ and MPs. It was found that $TiO_2$ exhibited a negative charge, due to the presence of Ti–OH and Ti–O– surface groups, under neutral and alkaline conditions [50,51]. Based on the quantity of BS, the CuO microparticles exhibited the same tendency as $TiO_2$ in rejecting MPs. The larger size of the CuO microparticles reduces their attachment to the functional groups of BS. SEM-EDS and AFM results confirmed that the amount of $TiO_2$ NPs on the fiber surface is higher than that of CuO. Therefore, $TiO_2$ has a more predominant influence on MPs rejection.

In comparing permeability and rejection performance, the PVDF-OH/BS-2/$TiO_2$ membrane demonstrated the best performance for the separation of MPs from water.

### 3.2.2. Oil-Water Emulsion Separation and Antifouling Test

The membranes were tested for evaluation of oil-in water emulsion. **Fig. 8** shows the permeability and the rejection of membranes confirmed by COD measurement. The membranes exhibited high permeability to emulsified oil-water, attributed to the presence of surfactant. Herein, Triton X-100 surfactant was used to prepare oil-water emulsion. Triton is non-ionic. Wang et al.[52] showed that with the presence of Triton X-

100 in the feed oil-water emulsion solution, superhydrophilic composite membrane was subject to pore wetting due to reduce the liquid entry pressure. As a result, PVDF membrane hydrophilicity enhances. It was observed that, adding surfactant to the saline feed solution reduced the surface tension of the feed [53]. This reduction in surface tension facilitated pore wetting due to the hydrophobic interactions between the non-polar tails of the surfactant and the hydrophobic PVDF membrane. Since the hydrophobic tails of the surfactant interact with the PVDF membrane surface, the hydrophilic heads are oriented towards the feed solution. This orientation attracts water molecules, thereby enhancing membrane permeability.

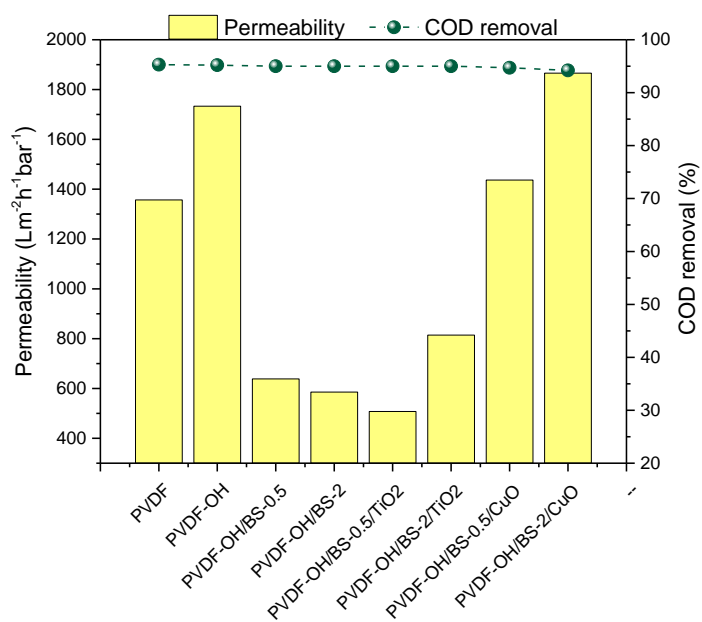

**Fig.8.** Oil-water emulsion separation performance of membranes.

**Table 2**. COD removal of oil-water emulsion at the first and 10$^{th}$ cycle.

| Sample | COD at 1$^{st}$ cycle (%) | COD at 10$^{th}$ cycle (%) |
| --- | --- | --- |
| PVDF | 95.3 | 95.1 |
| PVDF-OH | 95.2 | 95.2 |
| PVDF-OH/BS-0.5 | 95.0 | 94.9 |
| PVDF-OH/BS-2 | 95.0 | 94.8 |
| PVDF-OH/BS-0.5/TiO$_2$ | 95.0 | 95.1 |
| PVDF-OH/BS-2/TiO$_2$ | 95.0 | 95.0 |
| PVDF-OH/BS-0.5/CuO | 94.7 | 94.7 |
| PVDF-OH/BS-2/CuO | 94.2 | 93.9 |

Among all samples indicated in **Figure 8**, the alkali-treated PVDF-OH and CuO-modified membranes showed the highest permeability due to a higher hydrophilic surface. The smoother surface of the hydrophilic CuO compared to TiO$_2$ minimize fouling and offer less resistance to water flow, thereby enhancing the membrane's overall permeability and performance. Moreover, surfactant showed great stability and binding to the CuO NPs which might results in enhancement of membrane hydrophilicity [54]. On the other hand, the surfactant molecules' hydrophilic headgroups adsorbed onto the surfaces of TiO$_2$ particles, with their hydrophobic carbon chains oriented outward toward the feed solution. This interaction could result in the surfaces becoming more hydrophobic [55]. The higher surface roughness of TiO$_2$- modified membrane provides more sites for oil droplets to adhere, which can clog the pores and reduce flux. It must be noted that, before the emulsion separation test, DI water was used as the feed for each membrane to ensure stabilization and compaction. To maintain long-term performance and enhance membrane antifouling properties, the membranes were subjected to 10 consecutive cycles of use without undergoing any cleaning process between each cycle. Between each cycle, a pure water test was conducted to assess membrane fouling performance. **Figures S1-S4** (in the Supplementary Material) illustrate the oil-water and pure-water separation performance of the prepared membranes. **Table 2** presents the COD removal efficiency of the oil-water emulsion at both the initial and $10^{th}$ cycles. Based on the COD results, it can be concluded that all membranes demonstrated high oil removal efficiency (over 93%) even after the 10th cycle of the filtration process.

Clearly, the permeability of all membranes decreased after each cycle due to fouling. The pristine PVDF membrane became clogged after the $10^{th}$ cycle, resulting in almost zero permeability for both emulsion and subsequent DI-water separation (**Fig. S1**). Conversely, the one-step alkaline modification significantly improved overall permeability, maintaining over 1000 L/m²/h/bar even after the $10^{th}$ cycle. Compared to emulsion separation, DI-water separation exhibited lower permeability due to fouling.

After modification with biosurfactants, the membranes exhibited lower permeability compared to those treated with alkali (PVDF-OH) (**Figures S1 and S2**). Increased concentrations of the biosurfactant resulted in decreased permeability during emulsion separation. Comparing the $1^{st}$ and $10^{th}$ cycles of biosurfactant-modified membranes, both showed similar permeability at the end of the $1^{st}$ cycle. However, the membrane modified with a lower amount of biosurfactant (PVDF-OH/BS-0.5) demonstrated higher permeability than the one with a higher amount (PVDF-OH/BS-2) after the $10^{th}$ cycle. This suggests that using biosurfactants alone as a modifier is not suitable for emulsion separation. The interaction between oil droplets and the biosurfactant likely leads to oil droplets adhering to the membrane surface, causing fouling. It was found that Rhamnolipids exhibited efficient emulsifiers for vegetable oil and water, producing water-in-oil and oil-in-water emulsions [36].

COD results indicated that BS-modified membranes have similar oil removal efficiency to pristine and alkaline-treated PVDF (PVDF-OH). Even after the $10^{th}$ cycle, the membranes maintained the same removal efficiency as in the $1^{st}$ cycle.

The membranes were further modified with $TiO_2$ nanoparticles by incorporating various amounts of BS. Their emulsion and subsequent DI-water filtration tests were conducted as shown in **Fig. S3**. Membrane fouling occurs over time, and the performance of each cycle varies, sometimes being higher or lower than the previous cycle. This variation is due to the limited amount of feed used, constrained by the filter unit's capacity. For a more accurate evaluation, tests with larger amounts of feed solution are necessary. Based on the results in **Fig. S3**, it can be generally stated that the permeability performance of the $TiO_2$ modified membranes increased slightly compared to the biosurfactant (BS) modified membranes. The attachment of $TiO_2$ nanoparticles to the BS reduced the adhesion of oil droplets, thereby decreasing membrane fouling. The PVDF-OH/BS-2/$TiO_2$ membrane with a higher amount of BS showed slightly better permeability than the one with a lower amount of BS. Evidently, the presence of $TiO_2$ helps to reduce membrane fouling.

Lastly, CuO microparticles were used to modify the surface of the BS-modified membranes to observe the influence of these particles on membrane permeability for the separation of emulsified oily wastewater. Previous experiments, as shown in **Fig. S4**, indicated that adding TiO2 to the membrane surface slightly enhanced permeability. However, the addition of CuO to the membrane surface significantly improved permeability compared to pristine, alkali-treated, BS-modified PVDF, and TiO2 membranes. As previously explained, the hydrophilic nature of CuO increases membrane hydrophilicity and reduces the attachment of oil droplets, resulting in a substantial reduction in membrane fouling. Increasing the amount of BS led to a greater attachment of CuO on the membrane surface, which resulted in an increase in permeability. On the other hand, the oil removal rate of the CuO-modified membranes decreased slightly, possibly due to the larger average and maximum pore sizes of the CuO-modified membranes.

Besides water permeability and rejection, membrane antifouling tests are crucial for determining the overall quality of a membrane. Membrane antifouling was examined via flux recovery rate (FRR) and decline in permeability rate (Rt) at $10^{th}$ cycle, as calculated using equations 4 and 5. The results are presented in Table 3. The highest FRR with lowest Rt indicates the highest antifouling property. The membrane permeate flux decreases due to fouling with oil. The unmodified, pristine membrane exhibited the poorest recovery performance. Among all the samples, membranes containing $TiO_2$ nanoparticles demonstrated the best antifouling properties, achieving a high rejection rate (~95%). The anatase form of $TiO_2$ is highly hydrophilic, making it suitable for environmental applications [56]. However, it can be considered that hydrophilicity was not the only significant factor influencing membrane performance. In our case, the membranes exhibited hydrophobic characteristics when dry, likely due to their high surface roughness confirmed by AFM. The nanoparticles create micro- and nano-scale textures on the membrane surface, which might trap air and reduce the contact area between the water droplet and the membrane surface. Moreover, the rough surface of $TiO_2$ can minimize the contact points for oil droplets, reducing overall fouling and decline in permeability, while increasing the flux recovery rate [57].

**Table 3.** Antifouling performance of the modified and unmodified membranes.

| Sample | FRR(%) | Rt (%) |
|---|---|---|
| PVDF | 21.1 | 98.6 |
| PVDF-OH | 67.8 | 14.6 |
| PVDF-OH/BS-0.5 | 93.3 | 48.4 |
| PVDF-OH/BS-2 | 42.0 | 85.7 |
| PVDF-OH/BS-0.5/$TiO_2$ | 99.9 | 11.8 |
| PVDF-OH/BS-2/$TiO_2$ | 99.9 | 26.3 |
| PVDF-OH/BS-0.5/CuO | 60.9 | 31.6 |
| PVDF-OH/BS-2/CuO | 37.5 | 44.3 |

The particle leaching test to water is essential for assessing the attachment of micro- and nanoparticles to the membrane surface after modification. Leached particles can contaminate permeate water, making it unsafe for use, and alter the membrane's surface properties during the filtration process, leading to inconsistent performance. Furthermore, released particles can have adverse environmental impacts and must adhere to regulatory standards. Minimal leaching is critical for maintaining water quality, environmental safety, and cost efficiency in membrane applications. Leaching of nanoparticles from the membrane surface can degrade desired properties like antifouling performance and membrane lifespan [58]. Consequently, leaching tests for TiO2 and CuO particles were conducted using inductively coupled plasma mass spectrometry (ICP-MS) analysis. Table 4 presents the ICP-MS results of DI-water feed used in the leaching experiment. Several research articles indicate that leaching depends on the pH of the medium [59–61]. However, in this study, we maintained a neutral pH for both experiments involving particles and emulsion separation. The permeate was collected after reaching a total filtrate volume of 40 mL. It was found that the concentration of leached particles depends significantly on the amount of BS used to attach to the particles. This indicates that the quantity of modified particles depends on the amount of biosurfactant (BS). The results from ICP-MS corroborate the findings from SEM-EDS analysis.

**Table 4.** ICP-MS results of permeate after leaching experiments

| Sample | Concentration (μg/L) | wt. % from SEM-EDS analysis |
|---|---|---|
| PVDF-OH/BS-0.5/TiO$_2$ | 25.72 | 12.80 |
| PVDF-OH/BS-2/TiO$_2$ | 78.71 | 31.00 |
| PVDF-OH/BS-0.5/CuO | 62.28 | 0.20 |
| PVDF-OH/BS-2/CuO | 63.20 | 0.60 |

Kajau et al. [59] reported that the leaching of CuO from polyethersulfone (PES) membrane in water (under neutral conditions) is approximately 65 μg/L, consistent with our findings. In contrast, Li et al. [62] demonstrated a significantly higher leaching of TiO$_2$ nanoparticles from PES membrane (approximately 3.2 mg/L). In our study, the amount of leached TiO$_2$ nanoparticles depended on the quantity of biosurfactant (BS) used. The maximum concentration of TiO$_2$ in the water was measured as 78.71 μg/L, which is lower than reported in the literature.

*Comparison to current studies*

In comparison to current studies (**Table 5**), our research demonstrates a significant advancement in the modification of nanofibrous composite membranes for wastewater treatment. While previous studies have focused on individual modifications (**Table 5**), such as the use of TiO$_2$ or biosurfactants alone, our integrated approach combining alkaline treatment, biosurfactants, and nanoparticles (TiO$_2$ and CuO) has yielded superior performance in both permeability and contaminant rejection. Specifically, our TiO$_2$-modified membranes achieved higher microplastic (MP) rejection rates (99.99%) compared to those reported in earlier studies, which often did not exceed 90%. Additionally, the CuO-modified membranes exhibited enhanced permeability and antifouling properties, outperforming traditional PVDF membranes and even those modified with TiO$_2$ alone, highlighting the effectiveness of our multi-faceted modification strategy.

**Table 5.** Current studies on modified membranes for the separation of microplastics and oily wastewater: comparison with our research findings.

| Membrane type | Modification | Performance | Reference |
|---|---|---|---|
| Homoporous membranes fabrication using commercial PVDF | Initially adding amphiphilic surfactants, followed by a novelly-inserted air exposure progress to gently provide water vapor to induced a surface microscopic phase separation to achieve the growth of homopores. | Rejection rate over 97% (for 0.5μm PS MP), water permeability of 662 L m$^{-2}$ h$^{-1}$ bar$^{-1}$ | [63] |
| Crosslinked polyvinyl alcohol nanofibrous membrane | - | Rejection rate 77.3% +/- 1.4% at a pH of 6 (for 1μm PS MP) | [64] |
| polyurethane-based (PU) electrospun composite membrane | Graphene oxide-montmorillonite (GO -Mt) | Gravity-driven water flux of 793 L m$^{-2}$ h$^{-1}$ | [65] |
| PVDF nanofiber membrane | Modified with BS and TiO$_2$ | Rejection rate over 99.99% (for 0.5μm PS | This work |

| | | MP), water permeability> 5500 L m$^{-2}$ h$^{-1}$ bar$^{-1}$ | |
|---|---|---|---|
| PVDF ultrafiltration membrane | Modify with TiO2 | Initial flux of 400 L m$^{-2}$ h$^{-1}$ with a final flux of 200 L m$^{-2}$ h$^{-1}$ @ 2.5 h under 0.1 bar, oil rejection > 99.7% | [66] |
| PVDF ultrafiltration membrane | Modify with titanium dioxide/carbon nanotube (TiO2/CNT) nanocomposites | Max. flux of 1340 L m$^{-2}$ h$^{-1}$ at 0.3 MPa pressures, with oil rejection rate of 95.1-99.8%. | [67] |
| PVDF membrane | PVDF/poly(acrylic acid) (PAA) as the membrane matrix, TiO$_2$ nanoparticles as functional components and F127 as pore-forming agent | Pure water permeability reaches 1030 L m$^{-2}$ h$^{-1}$ and oil rejection rate> 93.4%. | [68] |
| PAN-Si nanofibers | Modifying with TiO$_2$ NPs | flux about 200,000 L m$^{-2}$ h$^{-1}$ bar$^{-1}$, separation of pump oil (1%) emulsion. | [69] |
| Hierarchical superflexible/superhydrophilic poly (vinylidene fluoride-co-hexa-fluoropropylene) (PVDF-HFP)/CuO-nanosheet membrane | Modifying with CuO by adding cupric acetate into electrospinning solution | Rejection rate over 99.89% (for >0.3μm PS MP), water flux 2360 ± 50 L m$^{-2}$ h$^{-1}$ @ 0.3 bar. | [70] |
| Super-hydrophobic/super-oleophilic carbon nanofibers (CNFs) | Chemical vapor deposition (CVD) with copper oxide (CuO) as the catalyst | Oil flux of 426 +/- 20 L m$^{-2}$ h$^{-1}$ with the water removal efficiency of 99.7% | [71] |
| Hierarchical TiO$_2$@CuO nanowire array-coated copper mesh membrane | TiO$_2$ and CuO | Water flux of 87.6 kL m$^{-1}$ h$^{-1}$ and separation efficiency with oil residue of only 12.4 mg L$^{-1}$ in the permeate. | [72] |
| Submerged membrane bioreactor (SMBR) | Combined with rhamnolipids surfactant | Removal efficiency of oil up to 90%, initial permeability of 220.75 L m$^{-2}$ h$^{-1}$ bar$^{-1}$ | [73] |
| PVDF nanofiber membrane | Modified with BS and CuO | Rejection rate over 94% (for oil-water emulsion), water permeability >1800 L m$^{-2}$ h$^{-1}$ bar$^{-1}$ | This work |

## Conclusion

The performed studies have demonstrated a successful approach to enhancing the separation performance of nanofibrous composite membranes for removing microplastics and oil from oil-water emulsions. Through the modification of PVDF polymeric nanofibers with BS, $TiO_2$, and CuO particles, the membranes demonstrated significant improvements in both permeability and rejection rates. The BS-modified membranes exhibited a substantial increase in water permeability, although this was accompanied by a decrease in particle rejection, likely due to BS release into the water which influence the membrane fouling due to reduction in hydrophilicity. The inclusion of $TiO_2$ nanoparticles provided the highest performance in microplastic rejection, attributed to the strong binding between $TiO_2$ and the microplastics, facilitated by their opposite charges. $TiO_2$ also enhanced antifouling properties and maintained high oil rejection rates (~95%) due to its hydrophilic nature and surface roughness.

Further modifications with CuO particles significantly improved membrane permeability, as CuO's hydrophilic properties reduced oil droplet attachment and fouling. Despite a slight decrease in oil removal efficiency, the CuO-modified membranes showed enhanced flux recovery and antifouling performance. The study also emphasized the importance of minimizing nanoparticle leaching to ensure water quality and environmental safety, with $TiO_2$ and CuO leaching levels maintained within acceptable limits.

It can be concluded that the modifications of PVDF membranes with BS, $TiO_2$, and CuO particles achieved a balance between high permeability and effective contaminant rejection, showcasing their potential for efficient wastewater treatment applications. This work demonstrates the feasibility of using modified nanofibrous composite membranes to address the challenges of microplastic and oil-water emulsion separation, providing a foundation for future advancements in membrane technology.

**ACKNOWLEDGEMENT:** Authors acknowledge the COST Action Horizon program and the COST Action PHOENIX: Protection, resilience, rehabilitation of damaged environment (CA19123), as well as the Research Infrastructure NanoEnviCz (Project No. LM2023066), supported by the Ministry of Education, Youth and Sports of the Czech Republic, and National Science Centre (Poland) OPUS grant (2019/33/B/NZ9/02774).